\documentclass[twoside,twocolumn,english,aps,prl,superscriptaddress,reprint]{revtex4-2}
\usepackage[T1]{fontenc}
\usepackage[latin9]{inputenc}
\usepackage{color}
\usepackage{babel}
\usepackage{amsmath}
\usepackage{amsthm}
\usepackage{graphicx}
\usepackage[unicode=true,pdfusetitle,
 bookmarks=true,bookmarksnumbered=false,bookmarksopen=false,
 breaklinks=false,pdfborder={0 0 1},backref=false,colorlinks=false]
 {hyperref}

\makeatletter

\usepackage{babel}
\usepackage{caption}

\usepackage{times}

\makeatother

\begin{document}
\global\long\def\ket#1{|#1\rangle}%
\global\long\def\bra#1{\langle#1|}%
\global\long\def\braket#1#2{\left\langle #1\middle|#2\right\rangle }%
\global\long\def\diff{\mathrm{d}}%
\global\long\def\inf{\infty}%
\global\long\def\pd#1#2{\frac{\partial#1}{\partial#2}}%
\global\long\def\Tr{\mathrm{Tr}}%

\title{Dynamical Phase Transitions of Information Flow in Random Quantum Circuits} 
\author{J.-Z. Zhuang} 
\email{zhuangjz21@mails.tsinghua.edu.cn} 
\affiliation{Center for Quantum Information, Institute for Interdisciplinary Information Sciences, Tsinghua University, Beijing 100084, PR China}
\author{Y.-K. Wu} 
\affiliation{Center for Quantum Information, Institute for Interdisciplinary Information Sciences, Tsinghua University, Beijing 100084, PR China} 
\affiliation{Hefei National Laboratory, Hefei 230088, PR China}
\author{L.-M. Duan} 
\email{lmduan@tsinghua.edu.cn} 
\affiliation{Center for Quantum Information, Institute for Interdisciplinary Information Sciences, Tsinghua University, Beijing 100084, PR China} 
\affiliation{Hefei National Laboratory, Hefei 230088, PR China}
\begin{abstract}
We study how the information flows in many-body dynamics governed
by random quantum circuits and discover a rich set of dynamical phase
transitions in this information flow. The phase transition points
and their critical exponents are established across Clifford and Haar
random circuits through finite-size scaling. The flow of both classical
and quantum information, measured respectively by Holevo and coherent
information, shows similar dynamical phase transition behaviors. We
investigate how the phase transitions depend on the initial location
of the information and the final probe region, and find ubiquitous
behaviors in these transitions, revealing interesting properties about
the information propagation and scrambling in this quantum many-body
model. Our work underscores rich behaviors of the information flow
in large systems with numerous phase transitions, thereby sheds new
light on the understanding of quantum many-body dynamics. 
\end{abstract}
\maketitle
\emph{Introduction.} Information flow in a quantum many-body system
usually accompanies the growth of quantum entanglement and drives
the system toward thermalization \citep{lewis-swanDynamicsQuantumInformation2019}.
Apart from being essential in understanding non-equilibrium many-body
physics, quantum dynamics about the information flow is also closely
related to black hole theory and quantum gravity through the AdS/CFT
correspondence \citep{harlowJerusalemLecturesBlack2016,qiDoesGravityCome2018}.
Under short-range interactions, the propagation of information is
limited in a light-cone structure governed by the Lieb-Robinson bound
\citep{liebFiniteGroupVelocity1972,rakovszkySignaturesInformationScrambling2019}.
On the other hand, in the long-time limit, quantum information scrambling
\citep{haydenBlackHolesMirrors2007,sekinoFastScramblers2008} will
occur for generic chaotic quantum systems, such that information initially
encoded in localized degrees of freedom will spread over the whole
system and cannot be recovered by local operations \citep{swingleUnscramblingPhysicsOutoftimeorder2018,shenkerBlackHolesButterfly2014,rakovszkySignaturesInformationScrambling2019}.
However, the detailed process between these two extreme cases is less
well-understood and involves rich phenomena like pre-thermalization
\citep{moriThermalizationPrethermalizationIsolated2018}, many-body
localization \citep{abaninColloquiumManybodyLocalization2019,nico-katzInformationtheoreticMemoryScaling2022},
and many-body scars \citep{turnerWeakErgodicityBreaking2018,yuanQuantumInformationScrambling2022}.

\begin{figure}
\includegraphics[width=1\columnwidth]{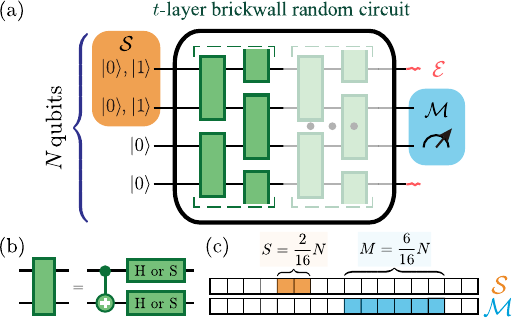}\caption{Model for probing information dynamics. (a) Information is encoded
into an $S$-qubit source in an $N$-qubit system with periodic boundary
condition. Then after $t$\textcolor{black}{{} layers of }brick-wall--structured\textcolor{black}{{}
random circuits}, we trace out the environment and retrieve the information
from the remaining $M$-qubit measurement subsystem. Each \textquotedblleft brick\textquotedblright{}
(green rectangle) represents a random operation between the two nearby
qubits. Here $N=4$ for illustration. (b) In each two-qubit random
operation, we first apply a CNOT gate. Then independently for each
qubit, we randomly apply a Hadamard or phase gate $\mathrm{diag}(1,e^{i\pi/2})$
with equal probability. (c) An example of source and measurement subsystem.
For ease of expression, they are consecutively selected according
to the $16$ equal segments of the system. \label{fig:Model-for-probing}}
\end{figure}

Here we study the information retrievable from a subsystem of an initially
locally encoded system, whose temporal derivative manifests the information
flow. We adopt the random quantum circuit ansatz as depicted in Fig.
\ref{fig:Model-for-probing}(a), which is widely used to capture universal
quantum dynamics in a chaotic system without being exposed to the
detailed Hamiltonian \citep{chandranSemiclassicalLimitManybody2015,vonkeyserlingkOperatorHydrodynamicsOTOCs2018,rakovszkyDiffusiveHydrodynamicsOutofTimeOrdered2018,khemaniOperatorSpreadingEmergence2018,changEvolutionEntanglementSpectra2019}.
By considering the information flow as a function of system parameters,
we uncover a spectrum of behaviors beyond the light-cone and scrambling
dynamics. Specifically, the information flow undergoes sudden shifts
and can be used to delineate phase boundaries as a function of time
and other system parameters. Similar to how order parameters switch
from zero to nonzero values across a phase boundary, it exhibits distinct
behaviors as the ratio of evolution time $t$ to the system size $N$
goes across the critical points in the thermodynamic limit $N\rightarrow\infty$,
thereby exhibiting dynamical phase transitions (DPTs). Note that there
are various notions of DPT in the literatures \citep{heylDynamicalQuantumPhase2018}.
The most widely used definition is based on the nonanalytical behavior
of the Loschmidt echo in closed many-body systems under Hamiltonian
evolution \citep{heylDynamicalQuantumPhase2013,zunkovicDynamicalQuantumPhase2018},
which is deeply connected to conventional partition functions. However,
this definition has no direct counterpart in random unitary circuits.
On the other hand, DPT is defined differently in open quantum many-body
systems \citep{atesDynamicalPhasesIntermittency2012,garrahanThermodynamicsQuantumJump2010}
and under the scenario of computational complexity \citep{deshpandeDynamicalPhaseTransitions2018}. 

Our key observation is the existence of dynamical phase transitions
and their universality across both classical and quantum information,
as well as within Clifford and Haar random circuits. We study their
physical meanings by quantifying the DPTs' positions and critical
exponents using finite-size scaling. We study primarily the Clifford
random circuits for the convenience of large-scale numerical simulation
\citep{gottesmanHeisenbergRepresentationQuantum1998,aaronsonImprovedSimulationStabilizer2004,fattalEntanglementStabilizerFormalism2004},
and generalization is verified for generic quantum circuit ansatz
\citep{nahumQuantumEntanglementGrowth2017,nahumOperatorSpreadingRandom2018}.
We also provide a general picture of information propagation that
applies to generic random circuit model setups. The discovery of such
rich phase transition behavior sheds new light on the understanding
of quantum many-body dynamics.

\emph{Dynamical Phase Transitions in Information Flow.} Consider an
$N$-qubit quantum system with periodic boundary condition, as shown
in Fig. \ref{fig:Model-for-probing}(a). We consecutively select $S$
qubits as the source of information $\mathcal{S}$ and $M$ qubits
as the measurement subsystem $\mathcal{M}$. We encode information
into $\mathcal{S}$, apply a random circuit $U$, trace out the complement
of $\mathcal{M}$ as the environment $\mathcal{E}$, and retrieve
the information from $\mathcal{M}$.

We first study the classical information dynamics in the quantum system,
and the quantum information will be discussed later. We encode $S$-bits
by preparing each qubit in $\mathcal{S}$ into $\ket 0$ or $\ket 1$
with equal probability. The rest of the qubits are initialized as
$\ket 0^{\otimes N-S}$. We denote the set of all the $2^{S}$ possible
initial states as $\left\{ \ket{\psi_{i}}\right\} _{i=1}^{2^{S}}$.
After the random circuit, the extractable information can be quantified
by the Holevo information 
\begin{equation}
H(U)=S_{vn}\left(\sum_{i}p_{i}\rho_{i}^{\mathcal{M}}\right)-\sum_{i}p_{i}S_{vn}\left(\rho_{i}^{\mathcal{M}}\right)\label{eq:Holevo-Definition}
\end{equation}
where $p_{i}=\frac{1}{2^{S}}$, $\rho_{i}^{\mathcal{M}}=\Tr_{\mathcal{E}}(U\ket{\psi_{i}}\bra{\psi_{i}}U^{\dagger})$
is the density matrix of $U\ket{\psi_{i}}$ in $\mathcal{M}$, $S_{vn}$
denotes the von Neumann entropy.

The random circuit comprises $t$ brick-wall layers, each corresponding
to a unit of abstract time. Despite of streched time scale, the structure
of information dynamics is uniform across different probability distributions
of the ``bricks'' over the Clifford group \citep{SupplementalMaterial_2023_InfoDynamics}.
As illustrated in Fig. \ref{fig:Model-for-probing}(b), we set each
brick as a CNOT gate followed by random single-qubit Clifford gates.
We denote $\mathcal{U}_{t}$ as the set of all possible $t$-layered
unitaries constructed in this way. 

With fixed $s=\frac{S}{N}$ and $m=\frac{M}{N}$, we study our system
under increasing system sizes $N$. We numerically calculate the time
evolution of average Holevo information $H(t)=\frac{1}{\left|\mathcal{U}_{t}\right|}\sum_{U\in\mathcal{U}_{t}}H(U)$
and normalize it by $h(t)=\frac{H(t)}{N}$. The averaged value $h(t)$
is sufficient to characterize each $h(U)$ for generic $U\in\mathcal{U}_{t}$
because, as we show in the Supplemental Material \citep{SupplementalMaterial_2023_InfoDynamics},
its variance over $\mathcal{U}_{t}$ vanishes in the large $N$ limit.
We also normalize the time by $\tau=\frac{t}{N}.$

\begin{figure}
\includegraphics[width=1\columnwidth]{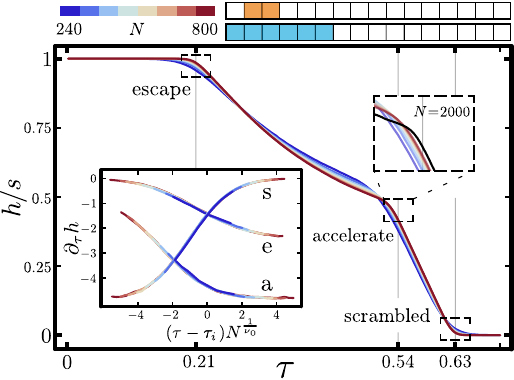}

\caption{Time evolution of average normalized Holevo information $h(\tau)$
under eight system sizes from $N=240$ (blue) to $N=800$ (red). We
fix $s\equiv\frac{S}{N}=\frac{2}{16}$ and the measurement subsystem
$m\equiv\frac{M}{N}=\frac{6}{16}$. At the three DPT\textcolor{blue}{{}
}points,\textcolor{blue}{{} }the curve becomes sharp as $N$ grows.
We denote them from left to right as the $\tau_{\mathrm{e}}$, $\tau_{\mathrm{a}}$,
and $\tau_{\mathrm{s}}$ point. For the $\tau_{\mathrm{a}}$ point,
we show an additional curve to illustrate its position. The inset
further demonstrates the transition by finite-size scaling of $\partial_{\tau}h$.
We find the critical exponent $\nu_{0}=1.25$ and scale the $\tau$-axis
near each of the three DPT points in the same way $\tau_{i}^{\prime}(\tau)=(\tau-\tau_{i})N^{\frac{1}{\nu_{0}}}$
where $i\in\{\mathrm{e},\mathrm{a},\mathrm{s}\}$. All of the eight
curves collapse. Each data point in the inset is obtained from over
$6\times10^{4}$ samples. \label{fig:Time-evolution-of-Holevo}}
\end{figure}

As an representative example, we place a $\frac{2N}{16}$-qubit source
inside a $\frac{6N}{16}$-qubit measurement subsystem. The information
dynamics $h(\tau)$ is shown in Fig. \ref{fig:Time-evolution-of-Holevo}.
In the limit of large $N$, three sharp turns of the curve can be
observed, indicating discontinuous $\partial_{\tau}h$ around the
three points. Further verification that they are DPT points and their
physical meanings will be discussed later. We denote them by their
$\tau$-axis position $\tau_{\mathrm{e}}$ (escape), $\tau_{\mathrm{a}}$
(accelerate), and $\tau_{\mathrm{s}}$ (scrambled). At early times
$\tau<\tau_{\mathrm{e}}$, $h(\tau)$ keeps its initial value $s$
because the light-cones starting from $\mathcal{S}$ are still inside
$\mathcal{M}$; at $\tau_{\mathrm{e}}$, information starts to decrease
by escaping through the left boundary of $\mathcal{M}$; when $\tau=\tau_{\mathrm{a}}$,
the rate of decreasing accelerates; after $\tau>\tau_{\mathrm{s}}$
, the system becomes scrambled $h(\tau)=0$, consistent with its infinite-time
limit \citep{zhuangPhasetransitionlikeBehaviorInformation2022}.

To further verify and analyze the critical behavior, we perform finite-size
scaling near each of the three DPT points $\tau_{i},i\in\{\mathrm{e},\mathrm{a},\mathrm{s}\}$.
The curves $\partial_{\tau}h$ of different system sizes $N$ collapse
when we scale the $\tau$-axis by the form $\tau_{i}^{\prime}(\tau)=(\tau-\tau_{i})N^{\frac{1}{\nu_{0}}}$
where we find the critical exponent $\nu_{0}$ to be equal for all
$i$, as shown in the inset of Fig. \ref{fig:Time-evolution-of-Holevo}.
Thus, around each $\tau_{i}$, we can express $\partial_{\tau}h$
of various $N$ as a same function of $\tau_{i}^{\prime}$. Then taking
the thermodynamic limit $N\rightarrow\infty$, we verify the non-analyticity
of information dynamics $\partial_{\tau}h\left(\tau_{i}-0\right)\neq\partial_{\tau}h\left(\tau_{i}+0\right)$.
As will be discussed later, the critical exponent $\nu_{0}$ is universal
across various model configurations.

\begin{figure}
\includegraphics[width=1\columnwidth]{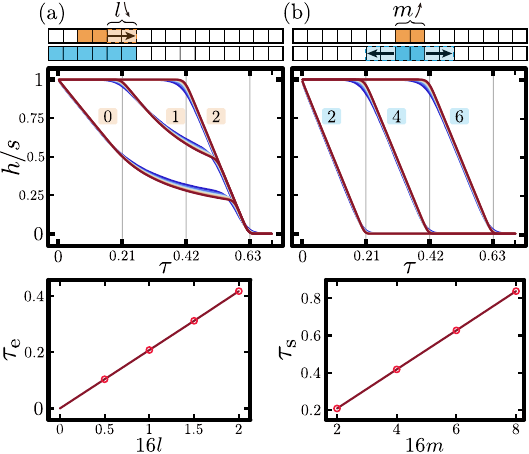}

\caption{Dynamics of $h(\tau)$ under different selections of the source $\mathcal{S}$
and the measurement subsystem $\mathcal{M}$. (a) We change the relative
position between $\mathcal{S}$ and $\mathcal{M}$. We keep $\mathcal{S}$
inside $\mathcal{M}$ and fix $s=\frac{2}{16}$, $m=\frac{6}{16}$.
Each group of curves is from various system sizes and labeled by the
corresponding relative position $16l$, where $l$ is the normalized
distance between the right boundaries of $\mathcal{S}$ and $\mathcal{M}$.
The escape point's position $\tau_{\mathrm{e}}$ is proportional to
$l$. The scrambled point $\tau_{\mathrm{s}}$ stays invariant. (b)
We change $\frac{2}{16}\protect\leq m\protect\leq\frac{1}{2}$ and
fix $\mathcal{S}$ in the middle of $\mathcal{M}$. Each group of
curves is labeled by $16m$. $\tau_{\mathrm{s}}$ is proportional
to $m$. For clarity, only part of the calculated $h(\tau)$ curves
are shown. \label{fig:Various-Selections-S-M}}
\end{figure}

\emph{DPTs' Positions and Physical Implications.} In order to determine
the physical meanings of the three DPT points, we study how their
positions can be determined by the selection of $\mathcal{S}$ and
$\mathcal{M}$. We begin by determining the $\tau$-axis positions
of the escape point $\tau_{\mathrm{e}}$ and the scrambled point $\tau_{\mathrm{s}}$,
before discussing the accelerate point $\tau_{\mathrm{a}}$.

With fixed $s$ and $m$, we move $\mathcal{S}$ from the middle to
the boundary of $\mathcal{M}$, as shown in Fig. \ref{fig:Various-Selections-S-M}(a).
By the periodic boundary condition, this only changes the relative
position of $\mathcal{S}$ and $\mathcal{M}$. We define the normalized
distance $l=\frac{L}{N}$ where $L$ is the minimal distance from
the qubits in $\mathcal{S}$ to the boundaries of $\mathcal{M}$.
When $l$ decreases, $\tau_{\text{e}}$ decreases linearly. When $l=0$,
the information can escape from $\mathcal{M}$ at the first circuit
layer, and the $\tau_{\text{e}}$ point disappears at $\tau=0$ as
expected. We have $\tau_{\text{e}}(l)=l/v_{\text{e}}$. The scrambled
point $\tau_{\text{s}}$, which marks the transition from $h>0$ to
$h=0$, is independent of $l$. Such independence holds for any arbitrary
selection of $\mathcal{S}$ \citep{SupplementalMaterial_2023_InfoDynamics},
as long as $h(\tau)$ is not exponentially so that $\tau_{\text{s}}$
does not vanish.

The scrambled point's position varies when $m$ changes, as shown
in Fig. \ref{fig:Various-Selections-S-M}(b) where we keep $\mathcal{S}$
fixed within $\mathcal{M}$. For $m<\frac{1}{2}$, we observe a linear
dependence $\tau_{\text{s}}(m)=\frac{m}{2}/v_{\text{s}}$. Combined
with invariant $\tau_{\text{s}}$ for arbitrary selections of $\mathcal{S}$,
this implies that no information can be retrieved from any consecutive
subsystem equal to or smaller than $m$ after $\tau_{\text{s}}$.
Non-consecutively selected subsystems of size $m$ are also scrambled.
One can see them scrambling faster by rearranging the qubits to make
them consecutive and the resulting circuit would contain longer-range
gates. We note that \textbf{$v_{\text{s}}$} can be directly connected
with the entanglement velocity $v_{\mathrm{E}}$ \citep{nahumOperatorSpreadingRandom2018}
, as demonstrated by the saturation of the entanglement entropy of $\mathcal{M}$ to its maximum value at $\tau_{\mathrm{s}}$.

We further found $v_{\mathrm{s}}=v_{\mathrm{e}}\equiv v_{\mathrm{I}}$
which we denote as the information velocity. The above analysis can
be summarized as 
\begin{equation}
\begin{cases}
\tau_{\mathrm{e}}(l)=l/v_{\mathrm{I}}\\
\tau_{\mathrm{s}}(m)=m/2v_{\mathrm{I}}
\end{cases}\label{eq:lightcone}
\end{equation}
which suggests a light-cone structure of information propagation underpinning
both the $\tau_{\mathrm{e}}$ and $\tau_{\mathrm{s}}$ points. $\tau_{\mathrm{e}}$
is the moment when light-cones emitted from the qubits in $\mathcal{S}$
reach the boundary of $\mathcal{M}$. Depending on whether they are
exiting or entering $\mathcal{M}$, information starts to either decrease
or increase. On the other hand, $\tau_{\mathrm{s}}$ is the moment
when $\mathcal{M}$ is entangled with $M$ qubits, reflected by the
light-cones emitted from $\mathcal{M}$ covering $2v_{\mathrm{I}}t_{\mathrm{s}}$
qubits outside of $\mathcal{M}$. To help understand the scrambling
condition $\frac{M}{M+2v_{\mathrm{I}}t_{\mathrm{s}}}\leq\frac{1}{2}$,
we note that this condition also applies when regarding the total
$M+2v_{\mathrm{I}}t_{\mathrm{s}}$ qubits as a maximally entangled
system \citep{zhuangPhasetransitionlikeBehaviorInformation2022}.
Such picture is also applicable both when $m>\frac{1}{2}$ and under
the open boundary condition \citep{SupplementalMaterial_2023_InfoDynamics}.

Though successful in predicting $\tau_{\text{e}}$ and $\tau_{\text{s}}$,
the light-cone picture cannot help understand the accelerate point
$\tau_{\text{a}}$. Specifically, the $\tau_{\text{a}}$ point is
not ``the time when light-cones start escaping from both ends of
$\mathcal{M}$''. For $l=0$ ($l=\frac{1}{16}$) in Fig. \ref{fig:Various-Selections-S-M}(a),
information can only reach the left boundary of $\mathcal{M}$ at
$\tau=\frac{4}{16}/v_{\mathrm{I}}$ ($\tau=\frac{3}{16}/v_{\mathrm{I}}$),
later than the actual $\tau_{\text{a}}$. We further show that $\tau_{\text{a}}$
has a nonlinear dependence on the model's length-scale \citep{SupplementalMaterial_2023_InfoDynamics},
suggesting that all linear light-cone understandings are insufficient.
Also, the accelerate point still exists when $\mathcal{S}$ has no
abrupt boundaries \citep{SupplementalMaterial_2023_InfoDynamics}. 

We have discussed above only $m<\frac{1}{2}$. For $m>\frac{1}{2}$,
the $\tau_{\mathrm{e}}$ and $\tau_{\mathrm{s}}$ points still exist
and are dominated by $v_{\mathrm{I}}$.\textcolor{blue}{{} }A major
difference is that the $\tau_{\mathrm{a}}$ point does not exist and,
after $\tau_{\text{s}}$, there are non-trivial dynamics followed
by another DPT which we denote as $\tau_{\mathrm{r}}$ (recover).
$h$ reaches minimum at $\tau_{\mathrm{s}}$ and saturates to its
non-zero infinite-time limit through $\tau_{\mathrm{r}}$ points.
More details can be found in the Supplemental Material \citep{SupplementalMaterial_2023_InfoDynamics}.

\emph{Dynamics of Quantum Information. }Coherent information quantifies
the reliably transmitted qubits through a noisy quantum channel. In
quantum communication, it characterizes the quantum channel capacity
when the encoding scheme is optimal \citep{barnumInformationTransmissionNoisy1998,holevoQuantumChannelsTheir2012}.
In quantum error correction, it upper bounds the number of qubits
that can be recovered \citep{schumacherQuantumDataProcessing1996}.
Using coherent information as a quantum counterpart of the Holevo
information, we compare classical and quantum information dynamics.

We now study the quantum information flow with similar encoding scheme
to that for the classical information. The only difference is that
the initial state of the source would be an ensemble $\rho^{\mathcal{S}}=\left(\frac{1}{2}I\right)^{\otimes S}$.
This is equivalent to mixing the pure states $\left\{ \ket{\psi_{i}}\right\} _{i=1}^{2^{S}}$
in the classical information model. Applying random circuit $U$ and
tracing out the environment $\mathcal{E}$ form a quantum channel.
The resulting coherent information $C^{\mathcal{M}}$ can be calculated
as \citep{schumacherQuantumDataProcessing1996,leditzkyDephrasureChannelSuperadditivity2018}
\begin{equation}
C^{\mathcal{M}}=S_{vn}(\rho^{\mathcal{M}})-S_{vn}(\rho^{\mathcal{E}})\label{eq:coherent-information}
\end{equation}
where $\rho^{\mathcal{M}}=\Tr_{\mathcal{E}}(U(\rho^{\mathcal{S}}\otimes\ket 0\bra 0^{\otimes N-S})U^{\dagger})$
is the density matrix of $\mathcal{M}$, and $\rho^{\mathcal{E}}$
is similarly defined by exchanging $\mathcal{E}$ and $\mathcal{M}$. 

Like what we have done to the Holevo information, we average the coherent
information over $\mathcal{U}_{t}$ and define $c=\frac{C^{\mathcal{M}}}{N}$.
We calculate $c(\tau)$ for various positions of $\mathcal{S}$ and
$\mathcal{M}$ with their sizes $s$ and $m$ fixed.

\begin{figure}
\includegraphics[width=1\columnwidth]{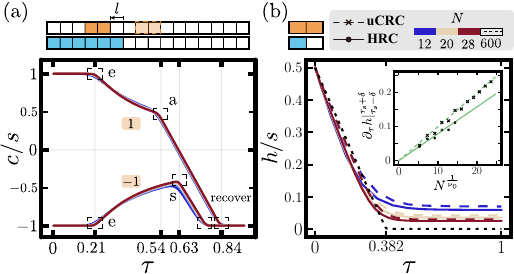}\caption{Universality of DPT points in quantum circuit ansatz. (a) Dynamics
of average normalized coherent information $c(\tau)$. We again specify
$s=\frac{2}{16}$ and $m=\frac{6}{16}$. $\mathcal{S}$ is inside
(outside) of $\mathcal{M}$ with the boundary distance $l=1$ (denoted
as $l=-1$). For $l=1$, up to three DPT points $\tau_{\mathrm{e}}$,
$\tau_{\mathrm{a}}$, and $\tau_{\mathrm{r}}$ can be observed. Additionally,
$c$ crosses from positive to negative at $\tau_{\mathrm{s}}$. For
$l=-1$, the $\tau_{\mathrm{a}}$ point does not exist and a DPT at
$\tau_{\mathrm{s}}$ appears. (b) Information dynamics in Haar random
circuits (HRC, solid line) and uniform sampling Clifford random circuits
(uCRC, dashed line). Here, we set the system size from $N=12$ (blue)
to $N=28$ (red) and fix $s=1,m=\frac{1}{2}$. $h(\tau)$ in both
systems exhibit similar behavior. The inset shows the scaling behavior
of $\left.\partial_{\tau}h\right|_{\tau_{\text{s}}-\delta}^{\tau_{\text{s}}+\delta}\propto N^{\frac{1}{\nu_{0}}}$
under both HRC and uCRC where the same critical exponent $\nu_{0}=1.25$
is applied. $\tau_{\text{s}}$ is from the thermodynamic limit of
uCRC and $\delta=0.012$ is a constant. \label{fig:Universality-DPTs}}
\end{figure}

As shown in Fig. \ref{fig:Universality-DPTs}(a), when $\mathcal{S}$
is inside (outside) of $\mathcal{M}$, $c$ is initialized at its
upper (lower) bound, indicating that all of the information are contained
in $\mathcal{M}$ (lost into $\mathcal{E}$). The escape point's position
$\tau_{\mathrm{e}}(l)=l/v_{\mathrm{I}}$ is the same as that in the
classical information model, verifying its physical meaning. For $\mathcal{S}$
inside $\mathcal{M}$, $c$ turns from positive to negative at $\tau_{\mathrm{s}}$,
indicating that the amount of remaining quantum information turns
to none. This is in agreement with the DPT point $\tau_{\mathrm{s}}$
in classical information dynamics. $c$ converges to its infinite-time
limit $-s$ through the last DPT. Consistent with the classical model
for $m>\frac{1}{2}$, we denote it as the $\tau_{\mathrm{r}}$ point. 

We can understand the above phenomena in the context of private classical
information transmission \citep{devetakPrivateClassicalCapacity2005,liPrivateCapacityQuantum2009,leditzkyDephrasureChannelSuperadditivity2018}.
We encode classical information by $\left\{ \ket{\psi_{i}}\right\} _{i=1}^{2^{S}}$
and give a penalty of $-1$ whenever one bit of information is leaked
to and recoverable from the environment. The result $C^{\mathcal{M}}=H^{\mathcal{M}}-H^{\mathcal{E}}$
is exactly the coherent information where $H^{\mathcal{M}}$ and $H^{\mathcal{E}}$
are the Holevo information in $\mathcal{M}$ and $\mathcal{E}$, respectively.
For $\mathcal{S}$ outside of $\mathcal{M}$, $H^{\mathcal{M}}$ stays
almost zero so that $C^{\mathcal{M}}\approx-H^{\mathcal{E}}$. $H^{\mathcal{E}}$
stays at maximum until starts decreasing at the $\tau_{\text{e}}$
point. At the scrambled point $\tau_{\text{s}}$, $H^{\mathcal{E}}$
reaches its minimum and starts recovering its value until saturation
at the $\tau_{\text{r}}$ point. With the size of $\mathcal{E}$ satisfying
$\frac{N-M}{N}>\frac{1+s}{2}$, $\mathcal{E}$ obtains all the $S$-bits
after $\tau_{\text{r}}$ while $\mathcal{M}$ acquires no information.

One can also regard the random circuit here as the encoding operation
in QEC \citep{brownShortRandomCircuits2013,choiQuantumErrorCorrection2020,gullansQuantumCodingLowDepth2021}.
Tracing $\mathcal{E}$ out would then correspond to the qubit loss
error, and $C^{\mathcal{M}}$ is the number of successfully preserved
logical qubits. In our result for $m<\frac{1}{2}$, the information
can never be perfectly recovered from $\mathcal{M}$ as long as $\mathcal{E}$
has non-zero overlap with $\mathcal{S}$, regardless of how deep the
encoding circuit is. On the other side, we can study the case when
$m>\frac{1}{2}$ by exchanging $\mathcal{E}$ and $\mathcal{M}$ so
that the qubits in $\mathcal{M}$ instead of $\mathcal{E}$ are lost.
From $C^{\mathcal{E}}=-C^{\mathcal{M}}$, all the encoded quantum
information can be recovered from $\mathcal{E}$ at the $\tau_{\text{r}}$
point. Our result gives the minimum and sufficient circuit depth for
a perfect QEC recovery.

\emph{DPTs in Haar random circuits.} To the best of our knowledge,
existing methods -- either analytical or numerical -- are incapable
of directly analyzing the information in large-scale Haar random circuit
(HRC) systems. We will demonstrate that the DPT structure of information
dynamics is universal across HRC and Clifford random circuits. Within
the precision achievable with current methods, the critical exponent
$\nu_{0}$ of DPTs is also universal. 

Specifically, we compare brick-wall circuits with two types of 2-qubit
bricks: those generated from Haar random unitary, and those generated
by uniformly sampling the Clifford group (uCRC), the latter being
a unitary 2-design \citep{zhuMultiqubitCliffordGroups2017}. For an
arbitrary pure initial state $\ket{\psi_{i}}$, the two circuits produce
identical average purity of $\mathcal{M}$ \citep{nahumOperatorSpreadingRandom2018}.
In order to reduce the finite-size drifts, we fix $s=1,m=\frac{1}{2}$
which gives a simple information dynamics containing only one DPT
point $\tau_{\text{s}}$. 

As shown in Fig. \ref{fig:Universality-DPTs}(b), $h(\tau)$ of HRC
for small system sizes $N\lesssim28$ behave similarly to uCRC. We
further demonstrate the universality of the critical exponent $\nu_{0}=1.25$
in the inset. For fixed $\delta$ satisfying $\delta N^{\frac{1}{\nu_{0}}}\ll1$,
$\left.\partial_{\tau}h\right|_{\tau_{\text{s}}-\delta}^{\tau_{\text{s}}+\delta}\equiv\partial_{\tau}h(\tau_{\text{s}}+\delta)-\partial_{\tau}h(\tau_{\text{s}}-\delta)$
of uCRC should be proportional to $N^{\frac{1}{\nu_{0}}}$. This can
be concluded from the collapsed $\partial_{\tau}h(\tau^{\prime})$
in the inset of Fig. \ref{fig:Time-evolution-of-Holevo} with non-zero
slope near $\tau^{\prime}=0$. When applying to HRC the same $\tau_{\text{s}}$
value from the thermodynamic limit of uCRC, the scaling behavior remains
consistent $\left.\partial_{\tau}h\right|_{\tau_{\text{s}}-\delta}^{\tau_{\text{s}}+\delta}\propto N^{\frac{1}{\nu_{0}}}$.
More details can be found in the Supplemental Material \citep{SupplementalMaterial_2023_InfoDynamics}.

\emph{Discussions}. In summary, we have studied the dynamical phase
transitions in information flow with universal behavior across random
unitary circuits. We identified up to four DPT points in both classical
and quantum information flow: escape, accelerate, scrambled, and recover.
We studied their dependence on the model configuration, uncovering
the light-cone structure of information propagation. The accelerate
and recover points show new stages of propagation other than ballistic
and scrambling behavior. The quantum circuit ansatz we focused on
already encompasses a broad range of quantum systems. The potential
for similar behavior in generic systems, especially those governed
by Hamiltonian dynamics, remains an area of great interest. Although
we discussed the DPTs only from the information perspective, we expect
similar DPT behavior in other important physical quantities. The discovery
of the DPTs shall thus shed new light on our understanding of generic
quantum many-body dynamics. 

\textbf{Acknowledgment:} We thank Z.-D. Liu and D. Yuan for discussions.
This work was supported by the Frontier Science Center for Quantum
Information of the Ministry of Education of China and the Tsinghua
University Initiative Scientific Research Program. The numerical calculations
in this study were carried out on the ORISE Supercomputer.

\end{document}


\global\long\def\ket#1{|#1\rangle}%
\global\long\def\bra#1{\langle#1|}%
\global\long\def\braket#1#2{\left\langle #1\middle|#2\right\rangle }%
\global\long\def\diff{\mathrm{d}}%
\global\long\def\inf{\infty}%
\global\long\def\pd#1#2{\frac{\partial#1}{\partial#2}}%
\global\long\def\Tr{\mathrm{Tr}}%

\title{Supplemental Material for "Dynamical phase transition of information flow in random quantum circuits"} 
\author{J.-Z. Zhuang} 
\email{zhuangjz21@mails.tsinghua.edu.cn} 
\affiliation{Center for Quantum Information, Institute for Interdisciplinary Information Sciences, Tsinghua University, Beijing 100084, PR China}
\author{Y.-K. Wu} 
\affiliation{Center for Quantum Information, Institute for Interdisciplinary Information Sciences, Tsinghua University, Beijing 100084, PR China} 
\affiliation{Hefei National Laboratory, Hefei 230088, PR China}
\author{L.-M. Duan} 
\email{lmduan@tsinghua.edu.cn} 
\affiliation{Center for Quantum Information, Institute for Interdisciplinary Information Sciences, Tsinghua University, Beijing 100084, PR China} 
\affiliation{Hefei National Laboratory, Hefei 230088, PR China}
\maketitle

\setcounter{equation}{0} \setcounter{figure}{0} \setcounter{table}{0}
\setcounter{page}{1} \makeatletter 
\global\long\def\theequation{S\arabic{equation}}%
 
\global\long\def\thefigure{S\arabic{figure}}%
 
\global\long\def\bibnumfmt#1{[S#1]}%
 
\global\long\def\citenumfont#1{S#1}%

We studied the dynamics of information in random quantum circuits
and discovered dynamical phase transitions. In the main text, we consecutively
select the information source $\mathcal{S}$ and the measurement subsystem
$\mathcal{M}$. Then we encode information into $\mathcal{S}$, apply
a brick-wall--structured\textcolor{black}{{} random circuit} $U\in\mathcal{U}_{t}$
and retrieve the information from $\mathcal{M}$. We probe the classical
information dynamics by the normalized Holevo information $h(U)=\frac{H(U)}{N}$.

Here, we give additional details of the classical information dynamics,
and discuss the univerality of DPTs in various random quantum circuit
models.

\section{Supplemental Details of Classical Information Dynamics}

\subsection{Vanishing Variance of Normalized Holevo Information Under Large $N$
Limit}

As mentioned in the main text, we only focused on the average value
of information $h(t)=\frac{1}{\left|\mathcal{U}_{t}\right|}\sum_{U\in\mathcal{U}_{t}}h(U)$
because it describes the behavior of each individual unitary in $\mathcal{U}_{t}$
under the thermodynamic limit. Here, we show that the variance of
$\left\{ h(U)|U\in\mathcal{U}_{t}\right\} $ converges to zero for
all $t$ when $N\rightarrow\inf$. We use the same model as in Fig.
2 in the main text and calculate the variance of $h(U)$ 
\begin{figure}[b]
\includegraphics[width=1\columnwidth]{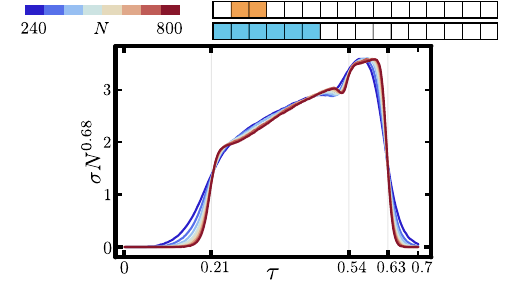}\caption{Standard variance $\sigma$ of the normalized Holevo information $h(U)$
over random circuits $U\in\mathcal{U}_{t}$. The curves for different
system sizes from $N=240$ (blue) to $N=800$ (red) approximately
collapse when $\sigma$ is scaled by $N^{0.68}$. The variance is
calculated by over $2\times10^{4}$ samples.\label{fig:Variation-of-Holevo}}
\end{figure}
\[
\sigma^{2}(\tau)=\text{Var}_{U\in\mathcal{U}_{t}}h(U)
\]

As shown in Fig. \ref{fig:Variation-of-Holevo}, we compute $\sigma$
for various system sizes and found $\sigma\sim N^{-0.68}$. Thus,
$\sigma\rightarrow0$ at the thermodynamic limit $N\rightarrow\inf$,
so $h(U)$ is the same for almost all $U\in\mathcal{U}_{t}$.

\subsection{Nonlinear Dependence of Accelerate Point $\tau_{\text{a}}$ on Length-scale}

\begin{figure}
\includegraphics[width=1\columnwidth]{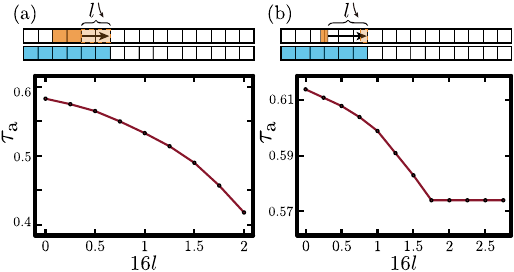}\caption{The accelerate point's position $\tau_{\text{a}}$ under different
selections of the source $\mathcal{S}$. $\mathcal{S}$ is moved from
the middle towards the boundary of $\mathcal{M}$. The normalized
distance from the right boundary of $\mathcal{S}$ to that of $\mathcal{M}$
satisfies $\frac{m-s}{2}\protect\geq l\protect\geq0$. We fix $m=\frac{6}{16}$.
(a) We set $s=\frac{2}{16}$. The accelerate point $\tau_{\text{a}}$
decreases nonlinearly when $l$ increases. (b) We set $s=\frac{0.5}{16}$.
The accelerate point $\tau_{\text{a}}$ decreases nonlinearly before
saturation to a fixed value. \label{fig:Accelerate-Point}}
\end{figure}

In the main text, we derived a light-cone picture by studying the
$\tau$-axis position of the escape point $\tau_{\mathrm{e}}$ and
the scrambled point $\tau_{\mathrm{s}}$. However, the accelerate
point $\tau_{\text{a}}$ and the recover point $\tau_{\text{r}}$
cannot be understood by the linear light-cone. 

Here, we study how the accelerate point $\tau_{\mathrm{a}}$ depends
on the selection of $\mathcal{S}$ and $\mathcal{M}$. As shown in
Fig. \ref{fig:Accelerate-Point}, when moving $\mathcal{S}$ from
the middle towards the boundary of $\mathcal{M}$, $\tau_{\text{a}}$
increases nonlinearly. Generally, the accelerate point occurs before
the scrambled point $\tau_{\text{a}}\leq\tau_{\text{s}}=\frac{m}{2v_{\mathrm{I}}}$.
For smaller $s$, we also oberserve a saturation behavior $\tau_{\text{a}}\geq\frac{m-s}{2v_{\mathrm{I}}}$.

\subsection{The Recover Point $\tau_{\text{r}}$}

\begin{figure}
\includegraphics[width=1\columnwidth]{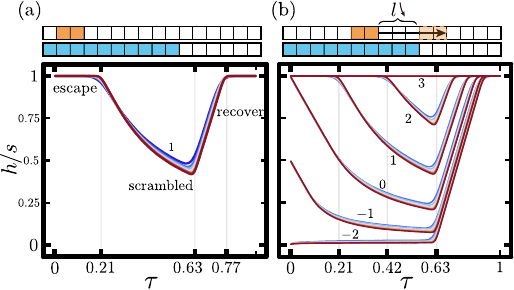}\caption{Dynamics of $h(\tau)$ under different selections of the source $\mathcal{S}$
and the measurement subsystem $\mathcal{M}$. We set $s=\frac{2}{16}$,
$m=\frac{10}{16}>\frac{1+s}{2}$. (a) The position of $\mathcal{S}$
and $\mathcal{M}$ are fixed corresponding to $l=\frac{1}{16}$ in
(b). (b) $\mathcal{S}$ is moved from the inside of $\mathcal{M}$
to the outside, and the corresponding group of curves can be found
from the top to bottom. Each group of curves is from various system
sizes and labeled by the corresponding relative position $16l$, where
$l$ is the normalized distance from the right boundary of $\mathcal{S}$
to that of $\mathcal{M}$. We set $l$ negative when $\mathcal{S}$
is not contained in $\mathcal{M}$. \label{fig:Recover-Point}}
\end{figure}

We discuss the information dynamics when $m>\frac{s+1}{2}$. Instead
of becoming scrambled as in $m<\frac{1}{2}$, the information recovers
its initial value at the long-time limit \citep{zhuangPhasetransitionlikeBehaviorInformation2022}.
As shown in Fig. \ref{fig:Recover-Point}(a), we set $s=\frac{2}{16}$,
$m=\frac{10}{16}$ and fix the position of $\mathcal{S}$ inside $\mathcal{M}$.
$h$ saturates to $s$ through the recover point $\tau_{\text{r}}$.
The whole propagation dynamics is as follows: $h=s$ for $0<\tau<\tau_{\text{e}}$.
When the information propagates across the boundary of $\mathcal{M}$
at $\tau_{\text{e}}$, information starts leaking out and $h$ starts
decreasing. Note that the minimal distance between boundaries of $\mathcal{S}$
and $\mathcal{M}$ is set to $l=\frac{1}{16}$, the same as that in
Fig. 2. As the result, $\tau_{\text{e}}$ is the same. At the scrambled
point, $h$ reaches its minimum. The position $\tau_{\text{s}}=\frac{1-m}{2}/v_{\mathrm{I}}$
is the same as in Fig. 2, where the measurement subsystem size is
$\frac{6}{16}=1-m$. $\mathcal{M}$ regains the information after
$\tau_{\text{s}}$ until the $\tau_{\mathrm{r}}$ (recover) point,
where $h$ saturates to its long-time limit. Combining with the $\tau_{\text{a}}$
point shown in the main text, we found in total four different DPT
points.

As shown in Fig. \ref{fig:Recover-Point}(b), we move $\mathcal{S}$
from the inside of $\mathcal{M}$ to the outside. Regardless of $\mathcal{S}$'s
position, the $\tau_{\text{s}}$ points share the same position $\tau_{\text{s}}=\frac{1-m}{2}/v_{\mathrm{I}}$.
In the main text, we intoduced the information light-cone picture.
For the complement of $\mathcal{M}$ with size $m^{\prime}\equiv1-m<\frac{1}{2}$,
$\tau_{\mathrm{s}}=\frac{m^{\prime}}{2}/v_{\mathrm{I}}$ is the moment
when $\mathcal{M}^{\prime}$ is entangled with $M^{\prime}$ qubits,
reflected by the light-cones emitted from $\mathcal{M}^{\prime}$
covering $2v_{\mathrm{I}}t_{\mathrm{s}}$ qubits outside of $\mathcal{M}^{\prime}$.
Here, the scrambled point of $\mathcal{M}$ shares the same $\tau$-axis
position. At $\tau_{\mathrm{s}}$, the number of qubits covered by
the light-cones ceases to increase and staturates to $N-M$. Correspondingly,
 the information stops decreasing. 

Though we found no method to infer the $\tau_{\text{r}}$ point's
position, there are some patterns that $\tau_{\text{r}}$ shares with
$\tau_{\text{a}}$. The $\tau_{\text{r}}$ points always occur later
than $\tau_{\text{s}}$. Moreover, as the relative position of $\mathcal{S}$
and $\mathcal{M}$ changes, the slope of the increasing $h(\tau)$
between $\tau_{\text{s}}$ and $\tau_{\text{r}}$ is invariant. The
slope of the decreasing $h(\tau)$ between $\tau_{\text{a}}$ and
$\tau_{\text{s}}$ in Fig. 3(a) is also invariant, but the two slopes
are not equal. 

\subsection{More Consecutive Selections of $\mathcal{S}$ and $\mathcal{M}$}

\begin{figure}
\includegraphics[width=1\columnwidth]{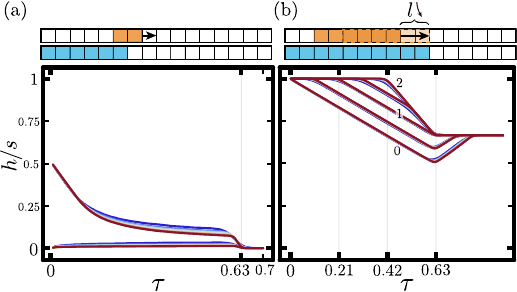}\caption{Dynamics of $h(\tau)$ under different selections of the source $\mathcal{S}$
and the measurement subsystem $\mathcal{M}$. (a) We set $s=\frac{2}{16}$
and $m=\frac{6}{16}$. $\mathcal{S}$ is moved out of $\mathcal{M}$.
(b) We set $s=\frac{6}{16}$ and $m=\frac{10}{16}$. The measurement
subsystem satisfies $\frac{1}{2}<m<\frac{1+s}{2}$ so that the information
$0<h(\tau\rightarrow\protect\inf)<s$. The normalized boundary-to-boundary
distance $l$ between $\mathcal{S}$ and $\mathcal{M}$ varies with
each step $\Delta l=0.5$. \label{fig:More-Consecutive-Selection}}
\end{figure}

In the main text, for $m<\frac{1}{2}$, we showed the evolution of
Holevo information when $\mathcal{S}$ is containted in $\mathcal{M}$.
Here, we study the case when $\mathcal{S}$ is not contained in $\mathcal{M}$.
As shown in Fig. \ref{fig:More-Consecutive-Selection}(a), when $\mathcal{S}$
is placed adjacent with $\mathcal{M}$, $h$ is initialized as $0$.
However, the information increases to a non-zero value for finite
$N$. It converges to zero when $N$ goes to infinity at a speed roughly
$\sim N^{-0.8}$, much slower than the exponential convergence speed
when scrambled $\tau>\tau_{\text{s}}$. This is in contrast with the
case when $\mathcal{S}$ is far away from $\mathcal{M}$, in which
$h(\tau)$ is exponentially small for all $\tau$ because the information
is scrambled before the light-cones reach $\mathcal{M}$. When half
of $\mathcal{S}$ is in $\mathcal{M}$ and the other half is not,
$h$ is initialized at $\frac{s}{2}$. For both positions of $\mathcal{S}$,
the $\tau_{\text{s}}$ point is invariant as in Fig. 2.

We also show the information dynamics when $\frac{1}{2}<m<\frac{s+1}{2}$.
See Fig. \ref{fig:More-Consecutive-Selection}(b), we set $s=\frac{6}{16}$
and $m=\frac{10}{16}$. $\mathcal{S}$ is placed inside $\mathcal{M}$
in the middle with the minimum distance between boundary $l=\frac{2}{16}$.
The infinite-time limit of $h(\tau)/s=(2m-1)/s=\frac{2}{3}$. As we
move $\mathcal{S}$ towards the boundary of $\mathcal{M}$, the DPT
types changes from $\tau_{\text{e}}$, $\tau_{\text{a}}$ and $\tau_{\text{s}}$
to $\tau_{\text{e}}$, $\tau_{\text{s}}$ and $\tau_{\text{r}}$.
We conclude that $\tau_{\text{a}}$ and $\tau_{\text{r}}$ does not
appear simultaneously for classical information propagation within
our model. Similar to $\tau_{\text{a}}$, $\tau_{\text{r}}$ also
has nonlinear dependence on the model's length-scale. Both of their
positions are hard to predict. We classify them into two types because,
as mentioned in the main text, they appear simultaneoulsy for quantum
information.

\section{Universality of DPTs in Random Circuit Ansatz}

Under the encoding-decoding protocal and the random quantum circuit
ansatz with brick-wall structure, we vary our model to verify the
existence of DPTs and the universality of their properties.

\subsection{Open Boundary Condition}

\begin{figure}
\centering

\includegraphics[width=1\columnwidth]{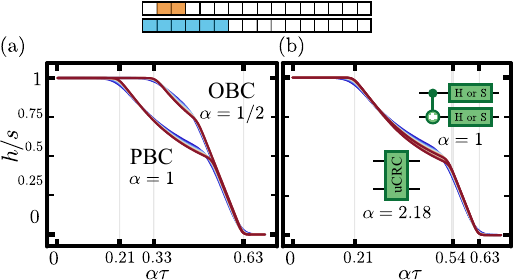}

\caption{We set $s=\frac{2}{16}$, $m=\frac{6}{16}$. (a) We compare the information
dynamics under periodic/open boundary condition. The $\tau$-axis
for the latter is rescaled linearly by $\alpha=\frac{1}{2}$. (b)
We compare the information dynamics with different random distributions
of Clifford unitary. Specifically, each two-qubit random operation
is generated by logic gates as in the main text or by sampling uniformly
over 2-qubit Clifford group. For the latter, the $\tau$-axis is rescaled
linearly by $\alpha=2.18$. Though $h(\tau)$ is different, the $\tau_{\text{e}}$
and $\tau_{\text{s}}$ points coincide. \label{fig:open-boundary}}
\end{figure}

With the same selection of $\mathcal{S}$ and $\mathcal{M}$, we compare
the dynamics of $h(\tau)$ under periodic and open boundary condition
(P/O BC), see Fig. \ref{fig:open-boundary}(a). Consistent with our
intuition, open boundaries block the propagation of information. The
information scrambles slower under OBC compared to PBC. However, this
does not change the structure of the information dynamics. As long
as $\mathcal{M}$ remains consecutive under OBC, all the DPT points
still exist with their physical meanings unchanged. Also, the critical
exponent for all the DPTs remains the same $\nu_{0}=1.25$.

We place $\mathcal{M}$ adjacent to the open boundary and fix $m=\frac{6}{16}$.
We further found that the position of the scrambled point under OBC
is exactly twice as that under PBC
\begin{align*}
\tau_{\text{s}}^{\text{OBC}} & =\frac{6}{16}/v_{\mathrm{I}}\\
\tau_{\text{s}}^{\text{PBC}} & =\frac{m}{2}/v_{\mathrm{I}}
\end{align*}

As shown in Fig. \ref{fig:open-boundary}(a), after rescaling the
$\tau$-axis for OBC by $\tau^{\prime}=\frac{1}{2}\tau$, the scrambled
point, as well as $h(\tau)$ between $\tau_{\text{a}}$ and $\tau_{\text{s}}$,
coincides for P/O BC. This is in agreement with the physical picture
of information propagation mentioned in the main text. Under PBC,
the light-cones from $\mathcal{M}$ propagates on both directions.
Here, the light-cones are blocked on the left side, and it takes a
doubled time for the region entangled with $\mathcal{M}$ to grow
and reach the size of $\mathcal{M}$. The halved increasing speed
of the affected region also corresponds to the halved decreasing speed
of information before $\tau_{\text{s}}$. 

We can also gain insights into the position of the escape point $\tau_{\text{e}}$
as follows. Under PBC, information escapes from the left side of $\mathcal{M}$
at $\tau_{\text{e}}^{\text{PBC}}=\frac{1}{16}/v_{\mathrm{I}}$. Under
OBC, information can only escape from the right boundary of $\mathcal{M}$,
which is three times the distance under PBC. In fact, $\tau_{\text{e}}^{\text{OBC}}\approx\frac{3}{16}/v_{\mathrm{I}}$
with $\tau_{\text{e}}^{\text{OBC}}$ relatively $5\%$ larger than
$3\tau_{\text{e}}^{\text{PBC}}$. The discrepancy is likely due to
the boundary effect because it extinguishes when $\mathcal{S}$ and
$\mathcal{M}$ are moved away from the boundary. In fact, when $\mathcal{M}$
is moved one unit right (figure not shown), the decreasing speed before
$\tau_{\text{s}}^{\text{OBC}}=\frac{5}{16}/v_{\mathrm{I}}$ is still
halved and there is a DPT point at $\tau_{\text{e}}^{\text{OBC}}=\frac{3}{16}/v_{\mathrm{I}}$
as expected.

\subsection{Other random Clifford circuits}

In the main text, each ``brick'' in the brick-wall--structured\textcolor{black}{{}
circuits represents} a CNOT gate followed by, for each qubit, a random
gate selected from the Hadamard gate and the phase gate $\mathrm{diag}(1,e^{i\pi/2})$
with equal probability. To verify the universality of DPTs over the
choice of Clifford random circuits, we generate each ``brick'' by
uniformly sampling the 2-qubit Clifford group (uCRC), as shown in
Fig. \ref{fig:open-boundary}(b). Though the $\tau_{\text{a}}$ points'
positions deviates with each other by $1\%$, the critical exponents
as well as the $\tau_{\text{e}}$ and $\tau_{\text{s}}$ positions
are identical when we rescale the $\tau$-axis $\tau^{\prime}=\alpha\tau$
by the same factor $\alpha$. Thus, the choice of Clifford random
circuits only affects the time scale of information propagation, rather
than its overall structure. 

\subsection{Haar random circuits}

\begin{figure}
\centering

\includegraphics[width=1\columnwidth]{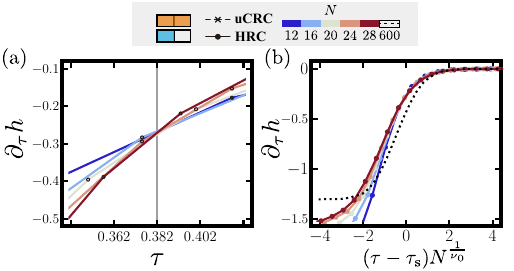}

\caption{Information dynamics in Haar random circuits (HRC, solid line). We
compare it with that in uniform sampling Clifford random circuits
(uCRC, dashed line). Here, we set the system size from $N=12$ (blue)
to $N=28$ (red) and fix $s=1,m=\frac{1}{2}$. (a) $\partial_{\tau}h$
for different $N$ overlap with each other at a same point. This determines
the DPT position at the crossing $\tau_{\text{s}}=0.382$. The same
value as the scrambled point of uCRC. (b) Finite-size scaling of $\partial_{\tau}h$.
The $\tau$ axis near $\tau_{\text{s}}$ is scaled by $\tau_{\text{s}}^{\prime}(\tau)=(\tau-\tau_{\text{s}})N^{\frac{1}{\nu_{0}}}$
where the critical exponent $\nu_{0}=1.25$. All of the fives curves
collapse. \label{fig:HaarClifford}}
\end{figure}

We next discuss the information dynamics in Haar random circuits (HRC).
To the best of our knowledge, existing methods -- either analytical
or numerical -- are incapable of directly analyzing the information
in large-scale HRC systems. We will provide evidences that the information
dynamics in HRC shares the same properties with that in uCRC.

Before numerical simulation results, we first undertake an analytical
examination of the connection between HRC and uCRC. For a cricuit
$U$ generated under the brick-wall structure, define $\rho_{i}^{\mathcal{M}}=\Tr_{\mathcal{E}}(U\ket{\psi_{i}}\bra{\psi_{i}}U^{\dagger})$
as the density matrix of $\mathcal{M}$. The average purity of $\mathcal{M}$
\[
\mathcal{P}=\Tr\left(\rho_{i}^{\mathcal{M}}\right)^{2}
\]
is identical \citep{nahumOperatorSpreadingRandom2018} for both circuits
under arbitrary circuit depth of $U$ and arbitrary pure initial state
$\ket{\psi_{i}}$. Note that in uCRC, each ``brick'' sampling from
the 2-qubit Clifford group forms a unitary $2$-design. Recall that
the Renyi-$\alpha$ entropy is defined by $S_{\alpha}=\frac{1}{1-\alpha}\log_{2}\Tr\left(\rho_{i}^{\mathcal{M}}\right)^{\alpha}$
whose $\alpha\rightarrow1$ limit equals the von Neumann entropy $S_{vn}\equiv S_{1}$.
As a special case of von Neumann entropy's generalization, $S_{2}$
shares the same mathematical structure and many properties with $S_{1}$
\citep{renyiMeasuresEntropyInformation1961}. The purity $\mathcal{P}$
is directly related with $S_{2}$ by 
\[
\mathcal{P}=2^{-S_{2}}
\]
This already ensures similar behavior of $S_{vn}\left(\rho_{i}^{\mathcal{M}}\right)$
contained in the second term of the Holevo information
\[
H(U)=S_{vn}\left(\sum_{i}p_{i}\rho_{i}^{\mathcal{M}}\right)-\sum_{i}p_{i}S_{vn}\left(\rho_{i}^{\mathcal{M}}\right)
\]
In summary, from the same average value of $\mathcal{P}$, the direct
relation from $\mathcal{P}$ to $S_{2}$ and the similarity between
$S_{1}$ and $S_{2}$, we conclude the similarity of Holevo information
dynamics between HRC and uCRC. 

For $h(\tau)$ under HRC, we can only simulate relatively small system
sizes $N\lesssim28$. In order to reduce the finite-size drifts, we
fix $s=1,m=\frac{1}{2}$ which gives a simple information dynamics
containing only one DPT point $\tau_{\text{s}}$, as illustrated in
Fig. 4(b) in the main text. To verify that the $\tau_{\text{s}}$
point's position is equal for HRC and uCRC, we perform the same finite-size
scaling for both systems. Fig. \ref{fig:HaarClifford}(a) shows $\partial_{\tau}h$
for various $N$. The DPT position can be determined by the crossing
$\tau_{\text{s}}=0.382$, consistent with uCRC. Fig. \ref{fig:HaarClifford}(b)
further demonstrates that the critical exponent in HRC is equal to
that in uCRC $\nu_{0}=1.25$. When we scale the $\tau$-axis by the
form $\tau_{i}^{\prime}(\tau)=(\tau-\tau_{i})N^{\frac{1}{\nu_{0}}}$,
$\partial_{\tau}h$ of different system sizes $N$ collapse near $\tau^{\prime}=0$.

\subsection{Independence of DPTs on the $\mathcal{S}$'s boundary }

\begin{figure}
\includegraphics[width=1\columnwidth]{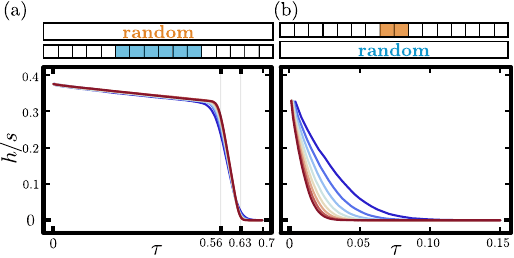}\caption{Dynamics of $h(\tau)$ under random selections of $\mathcal{S}$ or
$\mathcal{M}$. (a) $\mathcal{S}$ is chosen randomly while $\mathcal{M}$
is chosen consecutively. (b) $\mathcal{M}$ is chosen randomly while
$\mathcal{S}$ is chosen consecutively. \label{fig:Random-Selection}}
\end{figure}

In the main text, our consecutive selection of $\mathcal{S}$ creates
abrupt boundaries on both sides. Here, we consider how the DPTs depends
on the boundaries of $\mathcal{S}$. We pick out qubits from the whole
system randomly as $\mathcal{S}$ with the size still fixed by $s=\frac{2}{16}$.
Information now locate with various distance to the boundary of $\mathcal{M}$,
thus the time at which they start to escape from both ends of $\mathcal{M}$
distributes over a range of $\tau$ values. 

As shown in Fig. \ref{fig:Random-Selection}(a), when $\mathcal{S}$
has no abrupt boundries, the accelerate and the scrambled DPT point
still exists. The scrambled point's position is consistent with the
prediction $\tau_{\text{s}}(m)=\frac{m}{2}/v_{\mathrm{I}}$ in the
main text for $m=\frac{6}{16}$. The DPT points' existence is thus
independent on the boundary of information. Also note that the existence
of $\tau_{\text{a}}$ point demonstrates that it cannot be explained
directly by the light-cone viewpoint of the quantum information propagation. 

We also study the case when $\mathcal{M}$ is chosen randomly, as
shown in Fig. \ref{fig:Random-Selection}(b). At the thermodynamic
limit $N\rightarrow\inf$, the normalized time $\tau$ for the information
to scramble over the system tends to $\tau\rightarrow0$. This can
be understood by rearranging the qubits to make $\mathcal{M}$ consecutive
and the resulting circuit would contain long-range random interactions.
Thus, our results of the DPT points rely on consecutively selected
measurement subsystem. For a generic chaotic system with long-range
interactions, we expect it to be a fast-scrambler with no DPT exists
in its evolution. 

\subsection{DPTs in Quantum Information}

\begin{figure}
\includegraphics[width=1\columnwidth]{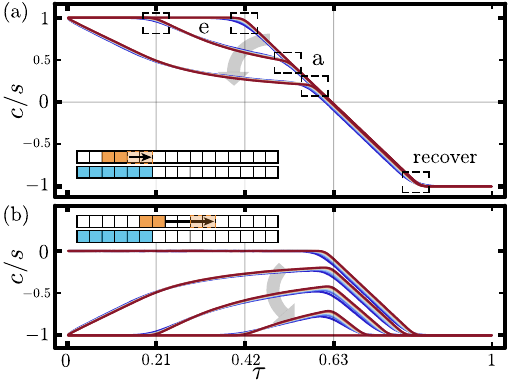}\caption{Time evolution of average normalized coherent information $c(\tau)$.
We again specify $s=\frac{2}{16}$ and $m=\frac{6}{16}$. $\mathcal{S}$
is moved from the middle of $\mathcal{M}$ to the outside, and the
corresponding group of curves can be found from top to bottom. (a)
$\mathcal{S}$ is contained in $\mathcal{M}$. Up to three DPT points
can be observed which we denote as the $\tau_{\text{e}}$, $\tau_{\text{a}}$
and $\tau_{\text{r}}$ point. $\tau_{\text{r}}$ is the same for various
selections of $\mathcal{S}$. Additionally, there is a transition
from positive to negative at the scrambled point $\tau_{\text{s}}$
which consists with the classical information model. (b) Part of or
all qubits in $\mathcal{S}$ are placed outside $\mathcal{M}$. The
$\tau_{\mathrm{a}}$ points do not exist and the DPTs at $\tau_{\text{s}}$
appears. \label{fig:Time-evolution-of-Coherent}}
\end{figure}

When $\mathcal{S}$ is contained in $\mathcal{M}$, as shown in Fig.
\ref{fig:Time-evolution-of-Coherent}(a), $c=s$ before the $\tau_{\text{e}}$
point. The position $\tau_{\text{e}}(l)=l/v_{\mathrm{I}}$ is consistent
with that in the classical information model. The speed of decay accelerates
also at the $\tau_{\text{a}}$ point. At time $\tau_{\text{s}}$,
$c$ turns from positive to negative, indicating that the amount of
remaining quantum information turns to none. This is in agreement
with the physical meaning of the scrambled point in classical information
dynamics. At the last DPT point, $c$ converges to its infinite-time
limit $-s$. Consistent with the classical model for $m>\frac{1}{2}$,
we denote it as the $\tau_{\text{r}}$ point. This can also be understood
from the perspective of the environment, in which the coherent information
$C^{\mathcal{E}}=-C^{\mathcal{M}}$ is fully recovered from its initially
negative value. The $\tau_{\text{r}}$ point is invariant when moving
$\mathcal{S}$ inside $\mathcal{M}$.

As shown in Fig. \ref{fig:Time-evolution-of-Coherent}(b), when half
of $\mathcal{S}$ is placed inside $\mathcal{M}$ and the other half
is outside, the coherent information stays at zero until a DPT around
$\tau_{\text{s}}$. When $\mathcal{S}$ is outside of $\mathcal{M}$,
$c$ is initialized at its lower bound $-s$, indicating that all
of the information are lost into $\mathcal{E}$. By definition, $L$
here is the distance from the left boundary of $\mathcal{S}$ to the
right boundary of $\mathcal{M}$. The $\tau_{\text{e}}$ point where
$c$ starts to increase has again $\tau_{\text{e}}(l)=l/v_{\mathrm{I}}$.
This verifies its physical meaning that the light-cone of information
reaches $\mathcal{M}$ from either outside or inside. $c$ increases
until reaching maximum at the scrambled point. After $\tau_{\text{s}}$,
$c$ decreases back to its lower bound until saturation through the
$\tau_{\text{r}}$ point.